\documentclass{PoS}

\title{Confining vs. conformal scenario for SU(2) with adjoint fermions. Gluonic observables.}

\ShortTitle{Confining vs. conformal scenario for SU(2) with adjoint fermions. Gluonic observables.}

\author{\speaker{Agostino Patella}\\
       CERN, Geneva, Switzerland\\
       E-mail: \email{agostino.patella@cern.ch}}

\author{Luigi Del Debbio\\
SUPA, School of Physics and Astronomy, University of Edinburgh, Scotland UK\\
E-mail: \email{luigi.del.debbio@ed.ac.uk}}
\author{Biagio Lucini\\
School of Physical Sciences, Swansea University, Wales UK\\
E-mail: \email{b.lucini@swansea.ac.uk}}
\author{Claudio Pica\\
CP$^3$-Origins \& IMADA, University of Southern Denmark, Odense, Denmark\\
E-mail: \email{pica@cp3.sdu.dk}}
\author{Antonio Rago\\
Department of Physics, Bergische Universit\"at Wuppertal, Germany\\
E-mail: \email{rago@physik.uni-wuppertal.de}}

\abstract{Walking technicolor is a mechanism for electroweak symmetry breaking without Higgs field. The Higgs mechanism is provided by chiral symmetry breaking in the technicolor theory. An essential ingredient is the vicinity to an IR fixed point, which could reconcile technicolor with the electroweak precision tests. SU($2$) gauge theory with two Dirac adjoint fermions has been proposed as a candidate for walking technicolor. Understanding whether this theory is confining or IR-conformal is a challenging problem, which can be addressed by means of numerical simulations. We have pointed out that a clean signal for the existence of an IR fixed point in this theory can be obtained by comparing the mesonic and gluonic sectors. We review some technical details of our calculations. Possible systematic errors are discussed.

\vspace{0.5cm}

\begin{flushright}
\parbox[t]{4cm}{
CERN-PH-TH/2010-253\\
CP3-Origins-2010-47\\
WUB/10-29\\
}
\end{flushright}
}

\FullConference{The XXVIII International Symposium on Lattice Filed Theory\\
                June 14-19,2010\\
                Villasimius, Sardinia Italy}

\begin{document}

\section{Introduction}

Very little is still know about gauge theories in the conformal window. Like QCD, IR-conformal gauge theories have a classical scale invariance which is anomalous at the quantum level and their quantum effective action is symmetric under chiral transformations. In QCD chiral symmetry is spontaneously broken and the trace anomaly defines the scale for the masses of all particles except the pion. Conversely in the conformal window chiral symmetry is expected to be intact and the trace anomaly is expected to decouple in the far infrared. Investigating the detailed dynamics of IR-conformal gauge theories has an intrinsic theoretical value.
In the last few years several numerical studies of gauge theories with many fundamental fermions or with fermions in higher representations have appeared in the literature (see~\cite{deldebbioPoS} and references therein). This renewed interest has been boosted by the proposal that theories inside or close to the conformal window could realize the technicolor mechanism for elecroweak symmetry breaking, while complying with the electroweak precision measurements (see for examples the reviews~\cite{Hill:2002ap, Lane:2002wv, Sannino:2009za, Piai:2010ma}).

Simulating an IR-conformal theory on the lattice is very challenging. Asymptotic scale invariance is explicitly broken by the finite volume, the nonzero mass for the fermions, the nonchiral discretization of the Dirac operator, and possibly the finite lattice spacing (if the IR fixed point has extra relevant directions). The twofold challenge is to develop analytical understanding in order to interpret the data correctly, and to have control of the systematics in simulations.

In this proceeding we review some of the results regarding the SU($2$) gauge theory with two Wilson-Dirac fermions in the adjoint representation (also known in literature as \textit{Minimal Walking Technicolor}). We have already presented and discussed all these results in~\cite{DelDebbio:2009fd, DelDebbio:2010hx, DelDebbio:2010hu}, which the reader should refer for details to. We will focus here in particular on the gluonic observables (Polyakov loop, glueballs, static potential, string tension), while we refer to the companion proceeding~\cite{picaPoS} for a short review of the mesonic observables.

\section{Numerical simulations and measured observables}

We have simulated the SU($2$) gauge theory with two Wilson-Dirac fermions in the adjoint representation. We used the HiRep code~\cite{DelDebbio:2008zf} which we have developed from scratch and extensively tested. All the data presented in this proceeding correspond to a single value of the coupling constant $\beta=2.25$. We consider here $16 \times 8^3$, $24 \times 12^3$ and $32 \times 16^3$ lattices. The long direction corresponds to Euclidean time with antiperiodic boundary conditions for fermions, while the short ones correspond to spatial directions with periodic boundary conditions. We simulated a wide range of masses corresponding to a PS triplet meson (the would-be pion in QCD) from about $0.2$ to $3$ in lattice units, or to a $(\textrm{string tension})^{1/2}$ from about $0.07$ to $0.4$. Of course PS masses above the cutoff cannot be taken too seriously. Nevertheless they are useful to illustrate some interesting features of this theory, like the spectrum hierarchy.

The simplest observable we measured is the distribution of the Polyakov loop for each direction. We point out that fermions in the adjoint representation do not break the $Z_2$ center symmetry, so the Polyakov loop is an order parameter for confinement.

Glueball masses have been measured with a variational technique (described in detail in~\cite{Lucini:2004my, Lucini:2004eq}), on a basis of operators obtained from small Wilson loops via a blocking-and-smearing procedure and projected over the irreducible representations of the cube symmetry group. As we will discuss, glueball masses are the most sensitive observables to finite-size errors. We managed to extract the $0^{++}$ glueball mass with some control on finite-size errors in a range between $0.5$ and $1.2$ in lattice units. Although we quote also the $2^{++}$ glueball mass, we have poor control over its finite-size errors.

The string tension has been extracted from the torelon masses, also computed with a variational technique. Once the torelon mass $E(T)$ (where $T$ is the size of the compact dimension) is computed, the string tension is extracted assuming a Nambu-Goto effective string (this choice is motivated in our paper~\cite{DelDebbio:2010hx}). The \textit{temporal} string tension is extracted from correlators of Polyakov loops winding in the time direction and separated by a spatial distance, while the \textit{spatial} string tension is extracted from correlators of Polyakov loops winding in a space direction and separated by a temporal distance. In a regime in which finite-size errors are negligible the temporal and spatial string tensions are expected to coincide.

\begin{figure}[ht]
\centering
\includegraphics*[width=.63\textwidth]{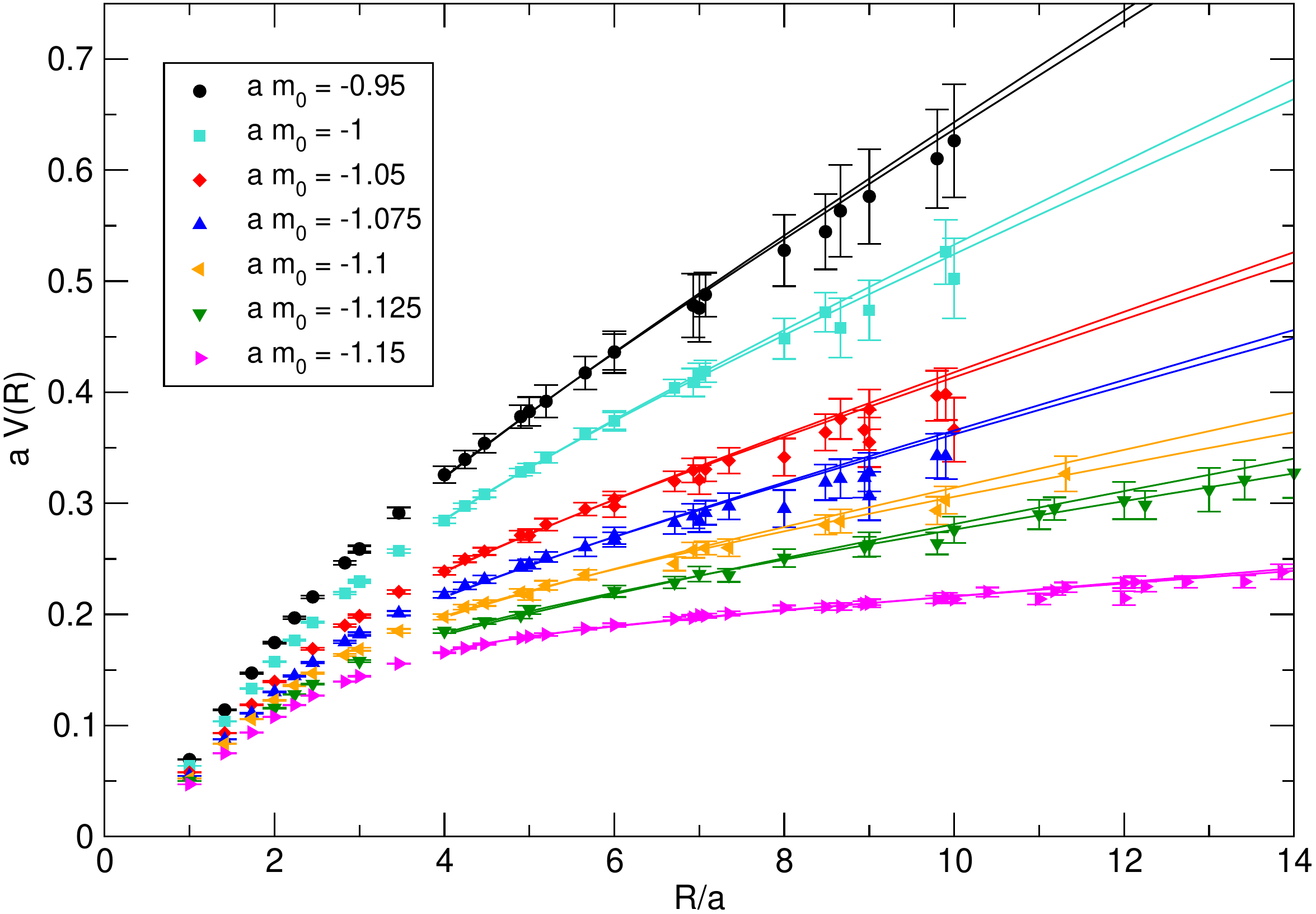}
\caption{Static potentials computed from Wilson loops. Superimposed is the function $V(R)=\sigma R + \mu + c/R$, where $\sigma$ is the string tension obtained from Polyakov loop correlators (the two curves correspond to $\sigma - \Delta \sigma$ and $\sigma + \Delta \sigma$), while $\mu$ and $c$ have been obtained with a fit in the region $R \ge 3a$.}
\label{fig:allpotentials}
\end{figure}

We also measured the static potential and force from HYP-smeared, on- and off-axis Wilson loops, using the Prony's method for isolating the ground state. The static potentials always show a clear linear asymptotic behavior, which signals confinement for every simulated nonzero fermionic mass (Fig.~\ref{fig:allpotentials}). Although it is quite hard to extract a string tension directly from the static force because of large statistical errors, the static force is always compatible with the string tension extracted from the Polyakov loop correlators.

\section{Main features and hints for conformality}

\begin{figure}[ht]
\centering
\includegraphics*[width=.63\textwidth]{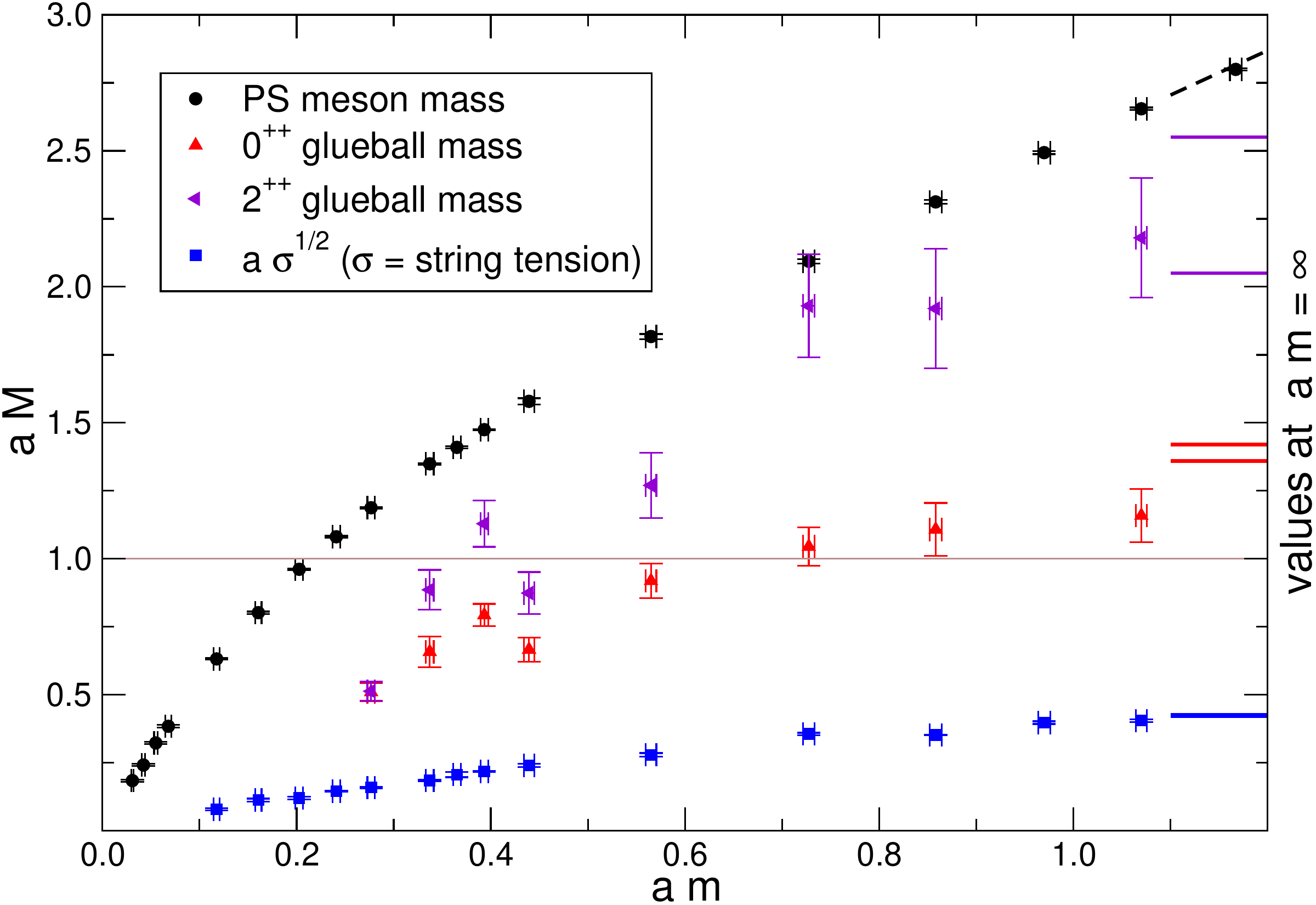}
\caption{The spectrum of the theory as a function of the PCAC mass
  $am$. The mass of the vector is not shown, since on the scale of the
  figure this state appears to be degenerate with the PS.}
\label{fig:spectrum_all}
\end{figure}

The plot in Fig.~\ref{fig:spectrum_all} shows the $(\textrm{string tension})^{1/2}$, the glueball masses and the PS triplet meson mass (the PS/V splitting can not be resolved on the scale of the plot) as functions of the PCAC mass. A striking feature is that the PS is heavier than the glueballs and the $(\textrm{string tension})^{1/2}$ for every simulated fermionic mass. At a first glance this could be dismissed as just the large-mass phase of a ($\chi$S broken or not) gauge theory. In fact asymptotically at large fermionic masses, the PS and V masses are expected to be almost degenerate and linear in the fermionic mass, and both these expectations matches our data. However in the very-large-mass regime, the glueball masses and the $(\textrm{string tension})^{1/2}$ would not depend on the fermionic mass. A moderately-large-mass regime could exists in which the PS and V masses are still almost degenerate, but the gluonic observables show a moderate dependence on the fermionic mass.

\begin{figure}[ht]
\centering
\includegraphics*[width=.63\textwidth]{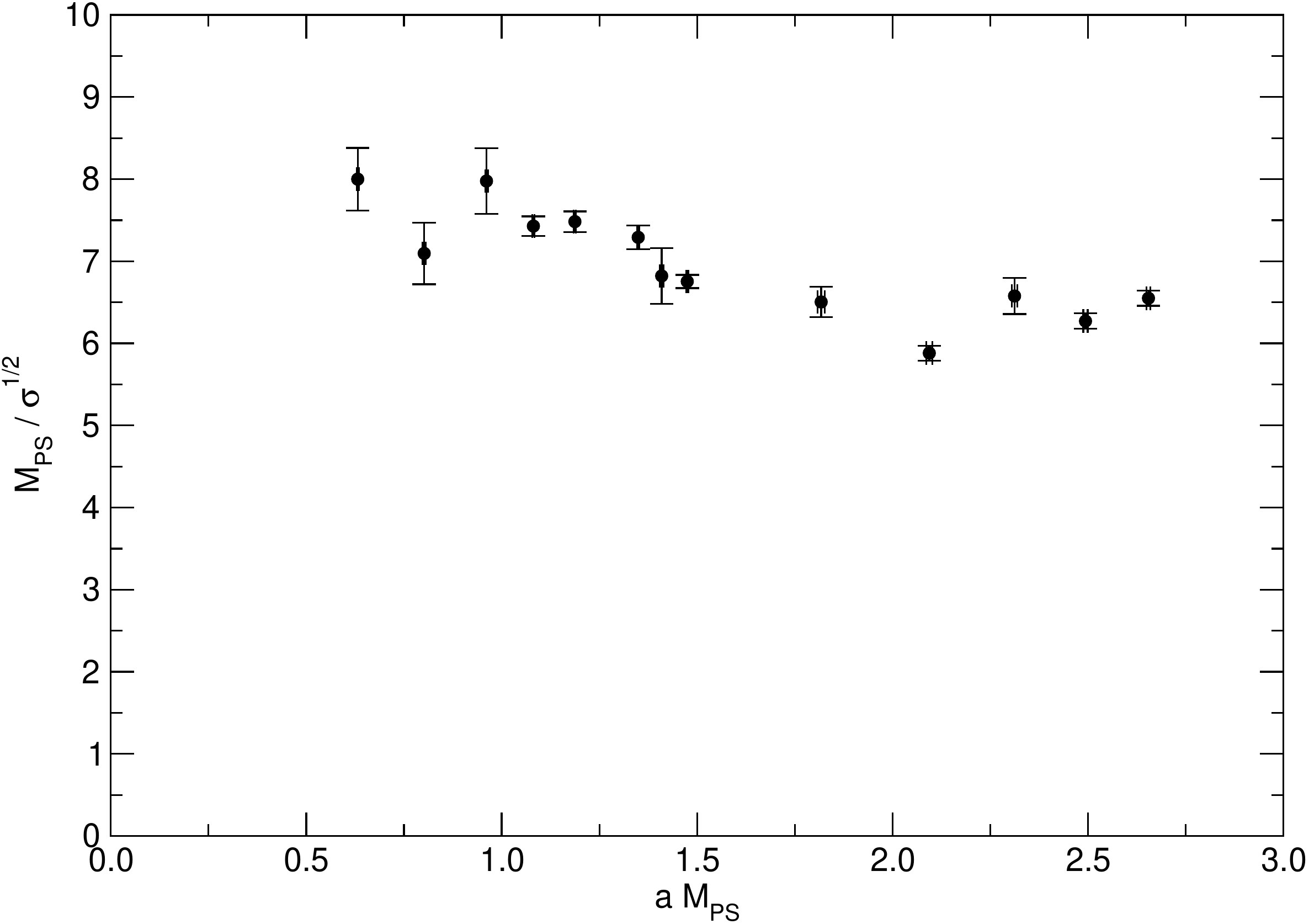}
\caption{The ratio $M_\mathrm{PS}/\sigma^{1/2}$ as a function of
  $M_\mathrm{PS}$.}
\label{fig:mps_over_sqrtsigma}
\end{figure}

\begin{figure}[ht]
\centering
\includegraphics*[width=.63\textwidth]{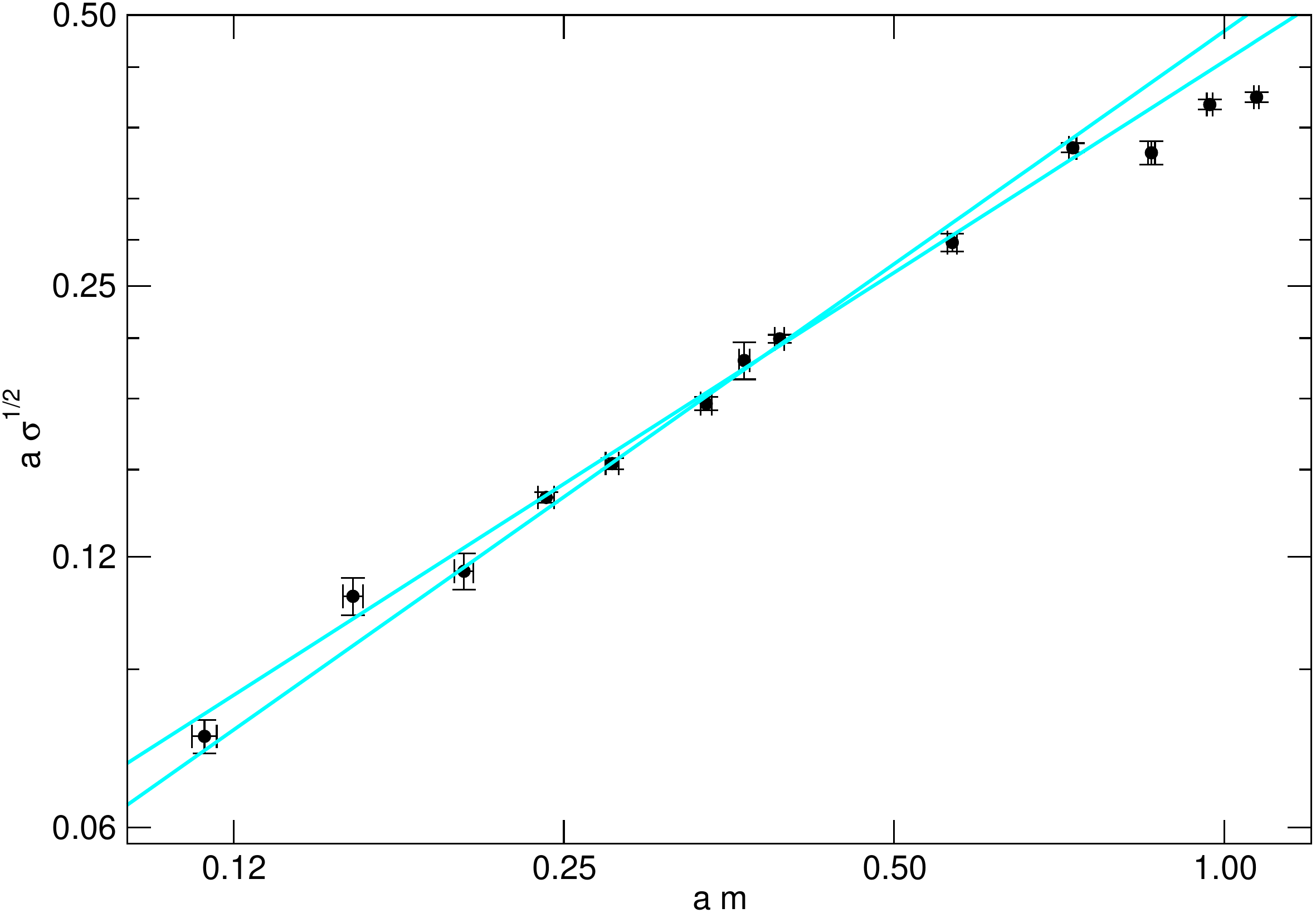}
\caption{$a\sigma^{1/2}$ as a function of the PCAC mass $am$. A fit of the data to
  a power law is also shown. In particular, the two lines
  represent the extremal values $\gamma_* = 0.16$ and $\gamma_* =
  0.28$.}
\label{fig:sqrtsigma_vs_mpcac}
\end{figure}

The scenario emerging from our data is dramatically different. While the PS mass drops of a factor $5$, the $(\textrm{string tension})^{1/2}$ also drops of the same factor. Their ratio, shown in Fig.~\ref{fig:mps_over_sqrtsigma}, stays approximatively constant (if anything, it increases at low masses) and between $6$ and $8$ in the whole simulated range. The large-mass regime is characterized by the $M_{PS}/\sqrt{\sigma}$ ratio linearly diverging in the fermionic mass, and is clearly not reached yet at the masses we simulate. On the other hand, in the chiral limit of a chirally broken theory the $M_{PS}/\sqrt{\sigma}$ ratio goes to zero, which also is very different from what we observe.
Our data are in fact compatible with the near-chiral behavior of an IR-conformal theory. In this case the masses of all particles are expected to go to zero in the chiral limit with the same power law in the PCAC mass~\cite{DelDebbio:2010ze,DelDebbio:2010jy}. Which in particular implies that ratios of masses go to a finite nonzero value in the chiral limit. We remind that we observed also such a behavior in the ratio of the PS and V masses~\cite{DelDebbio:2009fd, DelDebbio:2010hu}. We also fitted the string tension as a power of the PCAC mass. The fit works quite well in a wide range of masses, and gives an estimate for the anomalous dimension of the chiral condensate in the range between $0.16$ and $0.28$. In a log-log plot (Fig.~\ref{fig:sqrtsigma_vs_mpcac}), the power-law behavior is quite evident.\footnote{
Also a more exotic scenario can generate a finite nonzero value for $M_{PS}/\sqrt{\sigma}$ in the chiral limit: a confining (the string tension is different from zero) and chirally symmetric (the PS triplet meson stays massive) phase. However this is incompatible with the power-law behavior of the string tension.
}

\section{Some comments about finite-size effects}

In our simulations, gluonic observables are the most affected by finite-size errors, especially at lower fermionic masses. As an empirical observation, the finite-size errors on gluonic observables are related to the shape of the distribution of the Polyakov loop.

At infinite mass the fermions decouple and the theory is $SU(2)$ Yang-Mills with $\beta=2.25$, which has a critical radius (i.e. inverse critical temperature) between $3$ and $4$, and a $(\textrm{string tension})^{1/2}$ equal to $0.4231(25)$. Even on the $16 \times 8^3$ lattice, the string tension have finite-size effects not larger than few percents. In this situation the Polyakov loop distribution in every direction shows a clear single peak centered in zero, signaling confinement.
As the fermionic mass is decreased, the string tension decreases as well. If the volume (in lattice units) is kept fixed, the Polyakov loop develops a double-peak distribution below some \textit{critical mass}. Since the spatial directions are the short ones, this phenomenon happens in the distributions of spatial Polyakov loops first as the fermionic mass is decreased. We remark that this is not the deconfinement transition, and it can not be a phase transition at all, since the would-be center breaking happens in three (and not one) compactified directions.
In our simulations we always see a single-peak distribution for the temporal Polyakov loop, while the double peak in some spatial direction develops when $(\textrm{string tension})^{1/2}$ becomes smaller than the empirical value $1.65/L$. In particular this critical string tension decreases as the volume is increased.

When the double peak is generated, few things happen. The $0^{++}$ glueball mass stops being monotonous with the fermionic mass. Its finite-size errors seem to be not larger than the statistical ones above the critical mass (by direct comparison of two different volumes). Below the critical mass, the finite-size errors of glueball masses run immediately out of control. We discarded these points for the glueball masses. Above the critical mass, the spatial and temporal string tensions generally agree inside the statistical errors. Below the critical mass, the temporal string tension keeps decreasing and shows little finite-size effects (by direct comparison of two different volumes), while the spatial string tension flattens out. Again we kept only the points (some of them below the critical mass) for which we are reasonably confident that the finite-size errors are not larger than the statistical ones.

\section{Conclusions}

The presented investigation of the SU($2$) gauge theory with two Dirac fermions in the adjoint representation is based on the analysis of the spectrum, when the theory is deformed with a mass term for fermions. In the simulated range of masses, the theory is surely not in the chiral region of a chirally broken phase, since the PS triplet meson is not the lightest particle. The hierarchy of the spectrum can suggest that we are simulating in the large-mass regime of a QCD-like theory, but the stability of the $M_{PS}/\sqrt{\sigma}$ ratio together with the power-law behavior of the string tension would be hardly explained in this case. A more natural interpretation is in fact that we are simulating in the chiral region of an IR-conformal theory.
As we discussed in~\cite{DelDebbio:2009fd} (see also~\cite{Miransky:1998dh, Miransky:2010kb}), the particular hierarchy of the spectrum (a light glueball and heavy mesons) has a natural explanation, being the IR fixed point weakly coupled. In this case the coupling constant at the fixed point would be weak enough to generate a light glueball, but probably not enough to justify perturbation theory. This scenario is consistent with the smallness of the mass anomalous dimension.

The analysis presented above is based on the assumption that we are able to estimate the infinite-volume limit of the measured masses. We tried to convince the reader that, although we did not attempt any elaborate infinite-volume extrapolation, the finite-size errors are reasonably under control.

Finally we point out the limitations of our computation. First of all the described scenario could change going to lighter masses. Of course this objection always applies to numerical simulations. However if chirally symmetry breaking shows up at some lighter masses, this theory would become more (and not less) interesting. Showing chiral symmetry breaking at small fermionic masses, and conformal-like behavior at intermediate masses, it could represent a realization of walking. The main limitation of our computation remains the fact that we simulated at a single value for the coupling constant. Understanding whether the observed features survive to the continuum limit is of paramount importance, and is one of our main future goals.

\section*{Acknowledgments}

Simulations have been performed
on the BlueC supercomputer at Swansea University, on a Beowulf cluster partly funded by the Royal Society, and on the Horseshoe5 cluster at the University of Southern Denmark funded by a grant of the Danish Centre for Scientific Computing for the project \textit{Origin of Mass} 2008/2009.
Our work has been partially supported by STFC under contracts PP/E007228/1 and ST/G000506/1. BL is supported by the Royal Society (University Research Fellowship scheme), AP was supported by the EC (Research Infrastructure Action in FP7, project \textit{HadronPhysics2}). AR thanks the Deutsche Forschungsgemeinschaft for financial support.


\bibliography{patella}

\providecommand{\href}[2]{#2}\begingroup\raggedright\begin{thebibliography}{10}

\bibitem{deldebbioPoS}
L.~Del~Debbio, {\it {The conformal window on the lattice}},  {\em PoS} {\bf
  LATTICE2010} (2010) 004.

\bibitem{Hill:2002ap}
C.~T. Hill and E.~H. Simmons, {\it {Strong dynamics and electroweak symmetry
  breaking}},  {\em Phys. Rept.} {\bf 381} (2003) 235--402,
  [\href{http://xxx.lanl.gov/abs/hep-ph/0203079}{{\tt hep-ph/0203079}}].

\bibitem{Lane:2002wv}
K.~Lane, {\it {Two lectures on technicolor}},
  \href{http://xxx.lanl.gov/abs/hep-ph/0202255}{{\tt hep-ph/0202255}}.

\bibitem{Sannino:2009za}
F.~Sannino, {\it {Conformal Dynamics for TeV Physics and Cosmology}},  {\em
  Acta Phys. Polon.} {\bf B40} (2009) 3533--3743,
  [\href{http://xxx.lanl.gov/abs/0911.0931}{{\tt arXiv:0911.0931}}].

\bibitem{Piai:2010ma}
M.~Piai, {\it {Lectures on walking technicolor, holography and gauge/gravity
  dualities}},  \href{http://xxx.lanl.gov/abs/1004.0176}{{\tt
  arXiv:1004.0176}}.

\bibitem{DelDebbio:2009fd}
L.~Del~Debbio, B.~Lucini, A.~Patella, C.~Pica, and A.~Rago, {\it {Conformal vs
  confining scenario in SU(2) with adjoint fermions}},  {\em Phys. Rev.} {\bf
  D80} (2009) 074507, [\href{http://xxx.lanl.gov/abs/0907.3896}{{\tt
  arXiv:0907.3896}}].

\bibitem{DelDebbio:2010hx}
L.~Del~Debbio, B.~Lucini, A.~Patella, C.~Pica, and A.~Rago, {\it {The infrared
  dynamics of Minimal Walking Technicolor}},  {\em Phys. Rev.} {\bf D82} (2010)
  014510, [\href{http://xxx.lanl.gov/abs/1004.3206}{{\tt arXiv:1004.3206}}].

\bibitem{DelDebbio:2010hu}
L.~Del~Debbio, B.~Lucini, A.~Patella, C.~Pica, and A.~Rago, {\it {Mesonic
  spectroscopy of Minimal Walking Technicolor}},  {\em Phys. Rev.} {\bf D82}
  (2010) 014509, [\href{http://xxx.lanl.gov/abs/1004.3197}{{\tt
  arXiv:1004.3197}}].

\bibitem{picaPoS}
L.~Del~Debbio, B.~Lucini, A.~Patella, C.~Pica, and A.~Rago, {\it {Confining vs.
  conformal scenario for SU(2) with 2 adjoint fermions. Mesonic spectrum.}},
  {\em PoS} {\bf LATTICE2010} (2010) 069.

\bibitem{DelDebbio:2008zf}
L.~Del~Debbio, A.~Patella, and C.~Pica, {\it {Higher representations on the
  lattice: numerical simulations. SU(2) with adjoint fermions}},  {\em Phys.
  Rev.} {\bf D81} (2010) 094503, [\href{http://xxx.lanl.gov/abs/0805.2058}{{\tt
  arXiv:0805.2058}}].

\bibitem{Lucini:2004my}
B.~Lucini, M.~Teper, and U.~Wenger, {\it {Glueballs and k-strings in SU(N)
  gauge theories: Calculations with improved operators}},  {\em JHEP} {\bf 06}
  (2004) 012, [\href{http://xxx.lanl.gov/abs/hep-lat/0404008}{{\tt
  hep-lat/0404008}}].

\bibitem{Lucini:2004eq}
B.~Lucini, {\it {The large N limit from the lattice}},  {\em Few Body Syst.}
  {\bf 36} (2005) 161--166, [\href{http://xxx.lanl.gov/abs/hep-ph/0410016}{{\tt
  hep-ph/0410016}}].

\bibitem{DelDebbio:2010ze}
L.~Del~Debbio and R.~Zwicky, {\it {Hyperscaling relations in mass-deformed
  conformal gauge theories}},  {\em Phys. Rev.} {\bf D82} (2010) 014502,
  [\href{http://xxx.lanl.gov/abs/1005.2371}{{\tt arXiv:1005.2371}}].

\bibitem{DelDebbio:2010jy}
L.~Del~Debbio and R.~Zwicky, {\it {Scaling relations for the entire spectrum in
  mass-deformed conformal gauge theories}},
  \href{http://xxx.lanl.gov/abs/1009.2894}{{\tt arXiv:1009.2894}}.

\bibitem{Miransky:1998dh}
V.~A. Miransky, {\it {Dynamics in the conformal window in {QCD} like
  theories}},  {\em Phys. Rev.} {\bf D59} (1999) 105003,
  [\href{http://xxx.lanl.gov/abs/hep-ph/9812350}{{\tt hep-ph/9812350}}].

\bibitem{Miransky:2010kb}
V.~A. Miransky, {\it {Conformal phase transition in QCD like theories and
  beyond}},  \href{http://xxx.lanl.gov/abs/1004.2071}{{\tt arXiv:1004.2071}}.

\end{thebibliography}\endgroup

\end{document}